\documentclass[fleqn,twoside]{article}
\usepackage{espcrc2}
\usepackage{graphicx}
\usepackage[figuresright]{rotating}

\newcommand{\AmS}{{\protect\the\textfont2
  A\kern-.1667em\lower.5ex\hbox{M}\kern-.125emS}}
\title{Nuclear models and neutrino cross sections}
\author{Giampaolo Co\,'
        \address{Dip. di Fisica Universit\`a di Lecce
        and INFN sez. di Lecce, I-73100 Lecce, Italy}
        }
\begin{document}

\begin{abstract}
Merits and faults of the effective theory Random Phase Approximations
are discussed in the perspective of its use in the prediction of
neutrino-nucleus cross sections.
\vspace{1pc}
\end{abstract}
\maketitle
\begin{figure}[h]
\includegraphics[scale=0.45]
{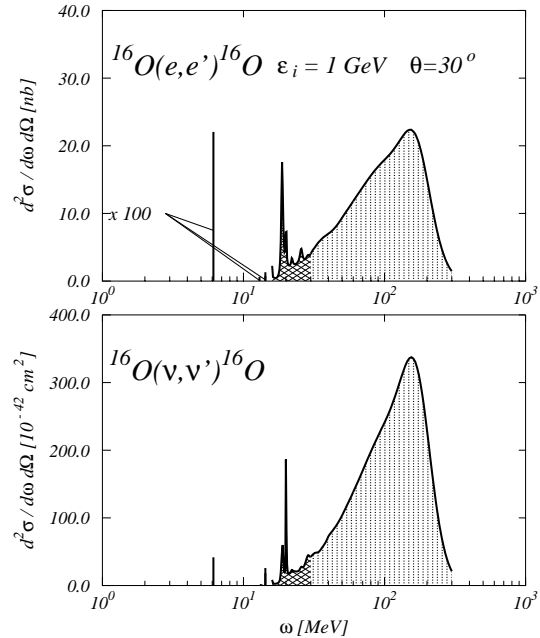}
\vspace{-1.5cm}
\caption{\small  
 Electron (upper panel) and neutrino (lower panel) cross sections.
}
\label{fig:resp}
\end{figure}

The Random Phase Approximation (RPA) is an effective theory aiming to
describe the excitation of many-body systems. In nuclear physics the
RPA has been applied with success over a wide range of excitation
energies. In Fig. \ref{fig:resp} we show inclusive (e,e') and
($\nu$,$\nu$') cross sections on $^{16}$O target nucleus, as a
function of the nuclear excitation energy. In both cases the incoming
energy of the lepton has been fixed at 1 GeV and the scattering angle
at 30$^o$.  In the figure, three different excitation regions are
emphasized. At few MeV of excitation energy there are discrete states,
from 15 up to 30 MeV there is the giant resonances excitation, and at
hundred of MeV the quasi-elastic peak.
%
%
%

The RPA describes the nuclear excited states as a linear combination
of particle-hole (ph) and hole-particle (hp) excitations
\begin{equation}
|\Psi_n>=\sum_{ph}(X_{ph}a^+_p a_h
 + Y_{ph}a^+_h a_p )|\Psi_o> \,.
\label{eq:psin}
\end{equation}
Aim of the theory is the evaluation of the $X_{ph}$ and
$Y_{ph}$ amplitudes for each excited state $|\Psi_n>$.
This is done by solving the secular equations
\[
\left(
\begin{array} {cc}
A & B \\ B^* & A^*
\end{array}
\right) 
\left(
\begin{array} {c}
X  \\ Y
\end{array}
\right) 
=  (E_n-E_o)
\left(
\begin{array} {c}
X  \\ -Y
\end{array}
\right) 
\,,
\]
where the coefficients of the matrix are expressed in terms of single
particle energies and wave functions as:
\begin{eqnarray}
\nonumber 
&~&A_{ph,p'h'} =
(\epsilon_p - \epsilon_h)\, \delta_{ph,p'h'}  + \\
&~& <ph'|V^{eff}|hp'> 
- <ph'| V^{eff}|p'h> \,,
\nonumber
\end{eqnarray}
and
\[ 
B_{ph,p'h'} = <pp'|V^{eff}|hh'> 
- <pp'|V^{eff}|h'h> 
\,.
\]

\begin{table} [ht]
\vskip 0.5 cm 
\begin{center}
\begin{tabular}{lccc}
\hline
 & $^{12}C$ & $^{12}N$ & $^{12}B$ \\
\hline
LM1 & 17.2 & 20.2 & 14.3 \\
LM2 & 18.8 & 21.7 & 15.9 \\
PP  & 16.7 & 19.6 & 13.8 \\
NuInt05 & 15.1 & 18.04 & 12.2 \\
exp & 15.1 & 17.3 & 13.4 \\
\hline
\end{tabular}
\end{center}
\caption 
{\small Energies, in MeV, of the isospin triplet 1$^+$ excited
states referred to the $^{12}$C ground state.
}
\label{tab:onep}
\end{table}

%
%
%
%
\begin{figure}[h]
\includegraphics[scale=0.45]
{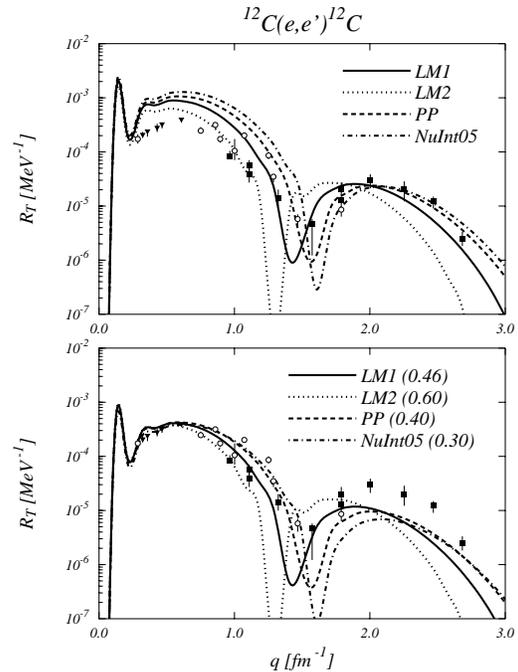}
\vspace{-1.0cm}
\caption{\small Magnetic form factors of the 1$^+$ state in 
  $^{12}$C calculated with the various interactions and compared with
  the experimental data. In the panel (b) the results of (a) have
  been multiplied by the quenching factors given in the figure in
  order to reproduce the data in the peak.
}
\label{fig:onep}
\end{figure}
%
%
%
%
%
\begin{figure}[h]
\includegraphics[scale=0.5]
{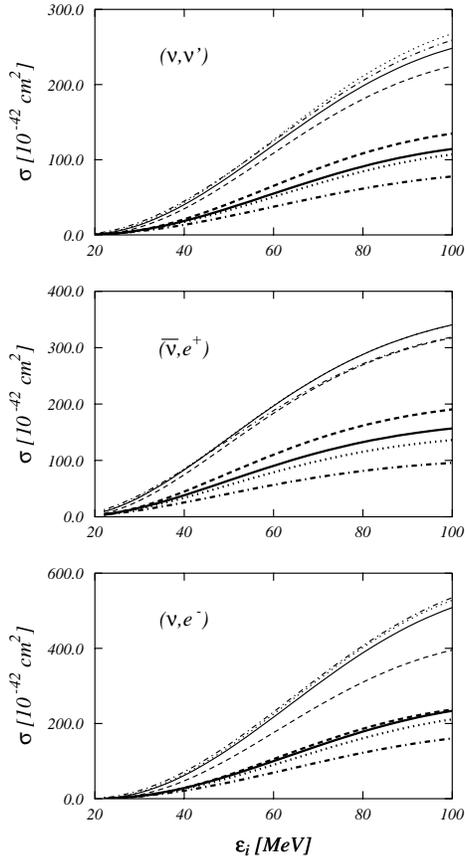}
\vspace{-1.0cm}
\caption{\small  
Total neutrino cross sections exciting the 
1$^+$ states of Tab. 
\protect\ref{tab:onep} as a function of the neutrino
energy. 
}
\label{fig:nuonep}
\end{figure}
%
%
%
%
\begin{figure}[h]
\includegraphics[angle=90, scale=0.5]
{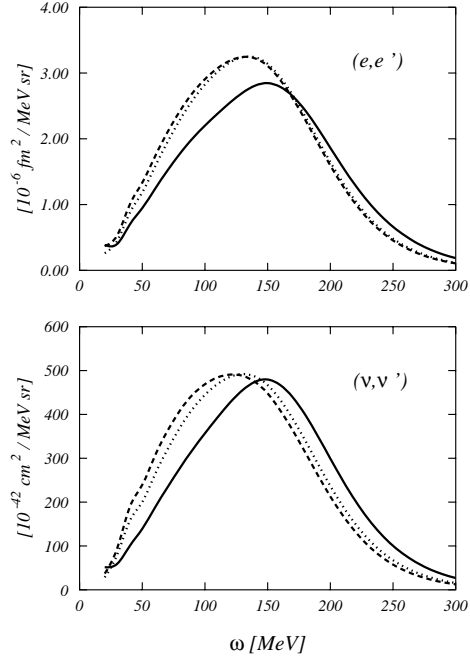}
\vspace{-1.0cm}
\caption{\small  
Electron and neutrino cross sections in the quasi-elastic region. 
The energy of the leptons is 1 GeV and the scattering angle 30$^o$.
The full lines have been obtained with RPA calculations using the LM1
interaction,  the dotted lines with the PP interaction. The dashed
lines show the mean-field results.
}
\label{fig:qerpa}
\end{figure}
\begin{center}
\begin{figure}[h]
\includegraphics[scale=0.38]
{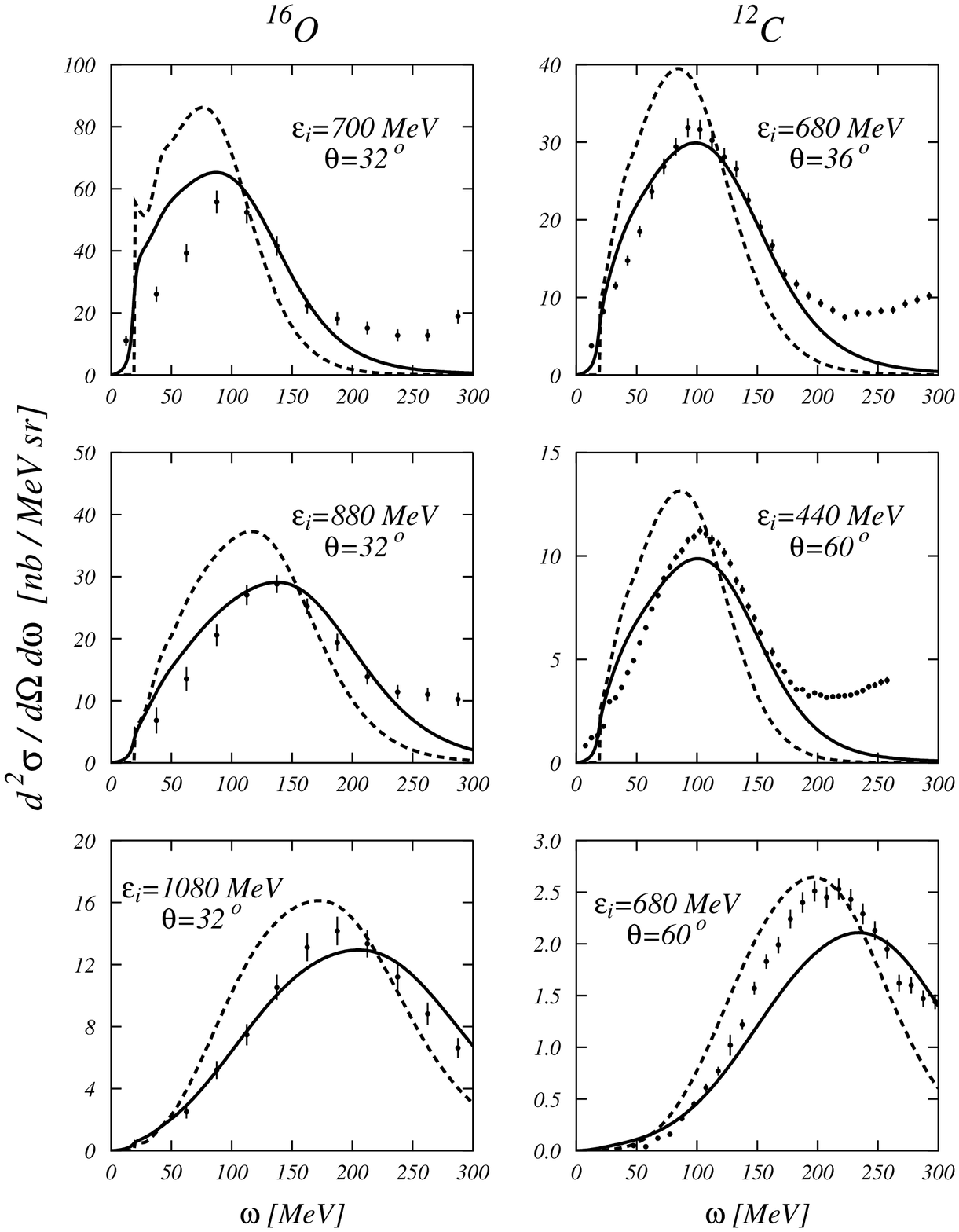}
\vspace{-1.0cm}
\caption{\small  
Electron scattering cross sections in the quasi-elastic region. The
mean-field results are shown by the dashed lines. The inclusion of the
FSI produces the full lines.
}
\label{fig:qfold}
\end{figure}
\end{center}
Single particle wave functions and energies are input of the theory.
In our calculations they have been generated by a Woods-Saxon
potential whose parameters have been fixed to reproduce the energies
of the levels close to the Fermi surface and the rms charge radii
\cite{bot05}.

The other input of theory is the effective interaction $V^{eff}$.
This is not the vacuum nucleon-nucleon interaction, but it is an
effective interaction in the medium, behaving well at short
internucleon distances.  We would like to point out the sensitivity of
the various RPA results on the choice of the effective interaction.
For this reason we used various effective interactions which have the
same dignity from first principles point of view.

We used two zero-range interactions whose parameters have been fixed
to reproduce muonic atom polarization isotope shifts in the $^{208}$Pb
region (LM1) \cite{rin78}, and spin responses in $^{12}$C
\cite{bak97}.  

We also used the polarization potential (PP) \cite{pin88} which is a
finite-range interaction whose parameters have been tuned to reproduce
some properties of nuclear matter. In addition we defined a new
interaction of zero-range type (NuInt05) whose parameters have been
fixed to reproduce the energy of the isovector 1$^+$ state in $^{12}$C
at 15.11 MeV.

In Tab. \ref{tab:onep} we compare the energies of the isovector 1$^+$
states in both charge conserving and charge exchange reactions with
the experimental values.  The two main issues we want to discuss in
the present report are already emerging from these results: the
sensitivity of the RPA results to the residual interaction and the
need of an explicit treatment of degrees of freedom beyond
one-particle one-hole excitations.  The uncertainty on the energies
calculated with the different interactions is of about 2 MeV.  Even
though the NuInt05 interaction reproduces the excitation energy of the
charge conserving 1$^+$ state, it is unable to reproduce the energies
of the charge exchange excitation. This indicates the limits of the
prediction power of the RPA theory. In the evaluation of some
observable, the use of effective, and phenomenological,
nucleon-nucleon interactions cannot substitute the explicit treatment
of many-particle many-hole excitations.

The above discussed issues are shown even better in Fig.
\ref{fig:onep} where the magnetic form factors of the 1$^+$ state in
$^{12}$C, calculated with the various interactions, 
are compared to the inelastic electron scattering data 
\cite{don68}. There is a large spreading of the theoretical curves. A
common feature is that all the results overestimate the data. This is
a well known problem of the RPA in the description of the magnetic
form factors of medium and heavy nuclei \cite{co90,mok00}.  It has
been argued that the explicit inclusion of many-particles many-holes
excitations can solve the problem \cite{kre80}. In a very crude and
phenomenological approach the various curves are multiplied by a
factor to reproduce the data in the maximum of the form factor.  The
values of these quenching factors are given in the panel (b) of Fig.
\ref{fig:onep}, where the renormalized results are shown.  Obviously,
the spreading between the various curves is reduced.

The consequences of these theoretical uncertainties on the neutrino
cross sections are presented in Fig. \ref{fig:nuonep} where the total
neutrino cross sections for the excitation of the three 1$^+$ states
of Tab. \ref{tab:onep} are shown as a function of the neutrino
energies.  The thinner upper lines are the bare RPA results, while the
lower curves, the thicker ones, have been obtained by multiplying the
RPA results with the quenching factors fixed in Fig. \ref{fig:onep}.

While in the electron excitation the use of quenching factors reduced
the spreading of the results, in the neutrino case this spreading has
increased.  This is a further indication of the fact that electrons
and neutrinos excite the same states in different manners.  In the
case of electrons the excitation is induced by a vector current while
neutrinos excitations are dominated by the axial vector current
\cite{bot05}.

The role of the effective interactions and of the many-particles
many-holes degrees of freedom in the giant resonance region, has been
thoroughly investigated in \cite{bot05} and we report here the main
results. The uncertainty on the total cross section is large for
neutrinos of 20-40 MeV. These uncertainties have heavy consequences on
the cross sections of low energy neutrinos such as supernova 
neutrinos and neutrinos coming from muon decay at rest.  When the
neutrino energy is above the 50 MeV, the results are rather
independent from the nucleon-nucleon interaction and indicate that the
inclusion of many-particles many-holes excitations reduces the RPA cross
sections by a 10-15\% factor.
 
We show in Fig.  \ref{fig:qerpa} an example of RPA effects in the
quasi-elastic region. Full and dotted lines have been calculated in
the continuum RPA framework by using the LM1 and the PP interaction
respectively. These results are compared to those obtained with a
mean-field model, i.e. by setting $V^{eff}$ to zero. The use of
zero-range interaction overestimates the RPA effects. They are
negligible when a finite-range interaction is used.  In the
quasi-elastic peak the the probe can resolve distances of about 0.5
fm, therefore zero-range interactions are not reliable.

Our calculations show that, in the quasi-elastic region, the RPA
effects are rather small. However, many-body effects beyond RPA are
not negligible, as it is shown in Fig. \ref{fig:qfold} where electron
scattering cross sections calculated within the mean-field model are
compared to data. In the quasi-elastic region, these complicated
many-body effects, which in the RPA language are described as
many-particles many-holes excitations, are usually called Final State
Interactions (FSI). 

Our treatment of the FSI assumes that they do not dependent on the
angular momentum ad the parity of the excitation.  Under this
assumption it is possible to correct RPA, or mean-field, responses for
the presence of FSI by folding them with a Lorentz function
\cite{co88}.  The parameters of this function are fixed by hadron
scattering data \cite{mah82}. We have shown that the inclusion of the
FSI reduces the quasi-elastic total neutrino cross sections by a
10-15\%  factor \cite{ble01}.

In summary, the effective theory RPA allows us to investigate
spectroscopic and dynamical properties of the nuclear excitations.
RPA calculations are necessary to produce giant resonances and
collective low-lying states.  The RPA results are however strongly
dependent on the effective nucleon-nucleon interaction used.
Interactions equivalent from the spectroscopic point of view, can
however produce very different excited states. The interactions we
have used give rather similar results for charge conserving natural
parity excitations. The situation for unnatural parity and
charge-exchange excitation is quite uncertain. In these cases it
emerges the necessity of including degrees of freedom beyond the RPA
assumptions since their effects cannot be simulated by readjusting the
parameters of the effective interaction. For neutrinos of energy
smaller than 100 MeV the neutrino-nucleus cross sections are still
very model dependent.

In the quasi-elastic region, which is dominated by single
particle dynamics, our results indicate that RPA effects are not
relevant. However the comparison with electron scattering data
shows the need of considering FSI, whose effects lowers the
total neutrino cross sections  by a 10-15\% factor.

\end{document}